\pdfoutput=1
\documentclass[12pt]{iopart}

\usepackage{graphicx}

\usepackage{epsfig}

\begin{document}
\title{Quark-Gluon Plasma - New Frontiers}

\author{Edward Shuryak}

\address{Department of Physics and Astronomy\\
University at Stony Brook,NY 11794 USA }

\ead{shuryak@tonic.physics.sunysb.edu}

\begin{abstract}
  As implied
by organizers,  this talk is not a conference summary but rather an 
outline of progress/challenges/``frontiers'' of  the theory.
Some fundamental questions addressed are:
 Why is sQGP such a
good liquid? Do we understand (de)confinement and 
what do we know about ``magnetic'' objects creating it? 
Can we understand
the AdS/CFT predictions, from the gauge theory side?
Can they be tested  experimentally?
Can AdS/CFT duality help us understand  rapid
equilibration/entropy production? Can we work out
a complete dynamical ``gravity dual'' to heavy ion collisions?
\end{abstract}

Heavy ion experiments continue to surprise  us.
At this conference 
 SPS experiments provided
 first dilepton flows from NA60,
and  ``conical flow'' from CERES. STAR and PHENIX at RHIC showed
a lot of new data on jet quenching
and jet correlations. Apparent absence of 
color Casimir factors in gluon/quark jets plus ``conical
flow'' suggests  {\em sound radiation} rather than gluonic
one. Heavy quark quenching is another part of this puzzle: and
we learned that up to a half of single electrons at the highest $p_t$
came from b quarks, we may soon learn the rate of $b$ quenching.
With coming LHC 
and the long-awaited low energy scan at RHIC,
 we will surely have more surprises ahead.

The situation in theory is still profoundly affected by
the paradigm shift occurred around 2003, to 
the strong-coupling regime.
 We are still in so-to-say non-equilibrium transition, as 
 huge amount of physics issues required  to be
learned.
Some came from other fields, including physics of {\em 
 strongly coupled QED plasmas} and
 trapped {\em  ultracold gases} with large scattering length.
{\em String theory} provided  a remarkable tool -- the AdS/CFT
correspondence -- which related heavy ions to the
the fascinating physics of strong gravity and {\em black
  holes}. Another important trend is that
{\em transport properties} of QGP and {\em non-equilibrium dynamics}
 came to the forefront: and for those the Euclidean approaches 
(lattice, instantons) we used before is much less suited
than for thermodynamics. All of it made the last 5 years
the time of unprecedented challenges.

\section{Pushing hydrodynamics beyond the O(10\%) level}

It is well known by now that hydro description of the QGP phase
supplemented by hadronic cascades  \cite{hydro+cascades}
 provides excellent
description of RHIC data. Radial and elliptic flow
of various secondaries, as a function of centrality,rapidity
or energy 
are reproduced till $p_t\sim 2 GeV$, which is 99\% of particles.
Contrary to predictions of some,
CuCu data match AuAu well, so  Cu is large
enough to be treated hydrodynamically. New hydro phenomenon
-- the ``conical flow'' \cite{conical} from jets -- got strong
conformation at QM08   from 3-particle
data from STAR and PHENIX. 

Thus sQGP is {\it the most
perfect liquid known}; before discuss why is it so, let us see
 how perfect is it? 
New round of studies last year focused on this issue, using
the so called second order formalism, which includes
viscosity and relaxation time parameters on top of ideal hydro. P. and
U.Romatschke
\cite{Romatschke:2007mq} were first, and they 
found that 
the best fits to $v_2(p_t)$ is at  $\eta/s\sim .03$,
 smaller than the famous AdS/CFT result \cite{Policastro:2001yc}
$\eta/s=1/4\pi$. Small viscosity effects in flow
were also found by Teaney
and Dusling
\cite{Dusling:2007gi} and by Chaudhuri (see his talk here):
with tensor correction at the freezout time dominating $v_2(p_t)$
as Teaney originally suggested.  D.Molnar (see his talk here)
have demonstrated  nice agreement between cascades and hydro
for $v_2(p_t)$, provided cross sections/viscosity
are appropriately tuned.
(Song and Heinz -- see  talk here --
 found for some reason larger viscous corrections to flow.)

Now, is the accuracy level really
 allows us to extract  $\eta/s$? 
The uncertainties in initial state deformation 
 are
 at the 10\% level (see Venugopalan's talk), thus comparable to the viscosity effect.
EoS  can probably be constrained better (lattice?). I think
uncertainties related to freezeout  -- not yet discussed
at all -- can also be reduced down to few percent 
 level, provided more efforts to understand  hadronic resonances/interactions
at the hadronic stage will be made. At the moment a safe statement 
  is  $\eta/s\sim 0.1$
and below .2 or so: while the exact value is still lacking
\footnote{Unfortunately
I am skeptical about magnitude of systematic 
errors of any lattice results for  
$\eta/s$
(such as \cite{Meyer:2007fc}): while the correlation functions
themselves are quite accurate, the spectral density
is obtained by rather arbitrary choice
between many excellent possible fits.
}.

Can viscosity   be even smaller, $\eta/s<1/4\pi$? 
In fact, as  Lublinsky and
myself \cite{Lublinsky:2007mm} discussed,  the  AdS/CFT 
gravity spectral densities predicts
that effective momentum-dependent 
viscosity $\eta(k)$ is $decreasing$ with momentum k,
from its famous value  $1/4\pi$ at k=0. We dont understand
its physics: but if so  this is very 
 important at very early stages, for most peripheral
collisions (thin almond), affecting the famous $v_2(centrality)$
curves on which hydro results are heavily based. 

\section{ A magnetic side of sQGP}
Long ago G.'t Hooft and Mandelstam \cite{t'Hooft-Mandelstamm} 
tried to explain 
confinement  by a ``dual superconductor'' made of Bose-condensed
 magnetically charged objects. 
 Seiberg and Witten \cite{SW}
have famously shown how it works in the \cal{N}=2 super Yang Mills
theory. Liao and myself \cite{Liao:2006ry} proposed a new view on
sQGP, based on electric-magnetic duality/competition, see Liao's talk
and also works by Zakharov et al \cite{Chernodub:2006gu}.

\begin{figure}[t]
  \includegraphics[width=0.48\textwidth]{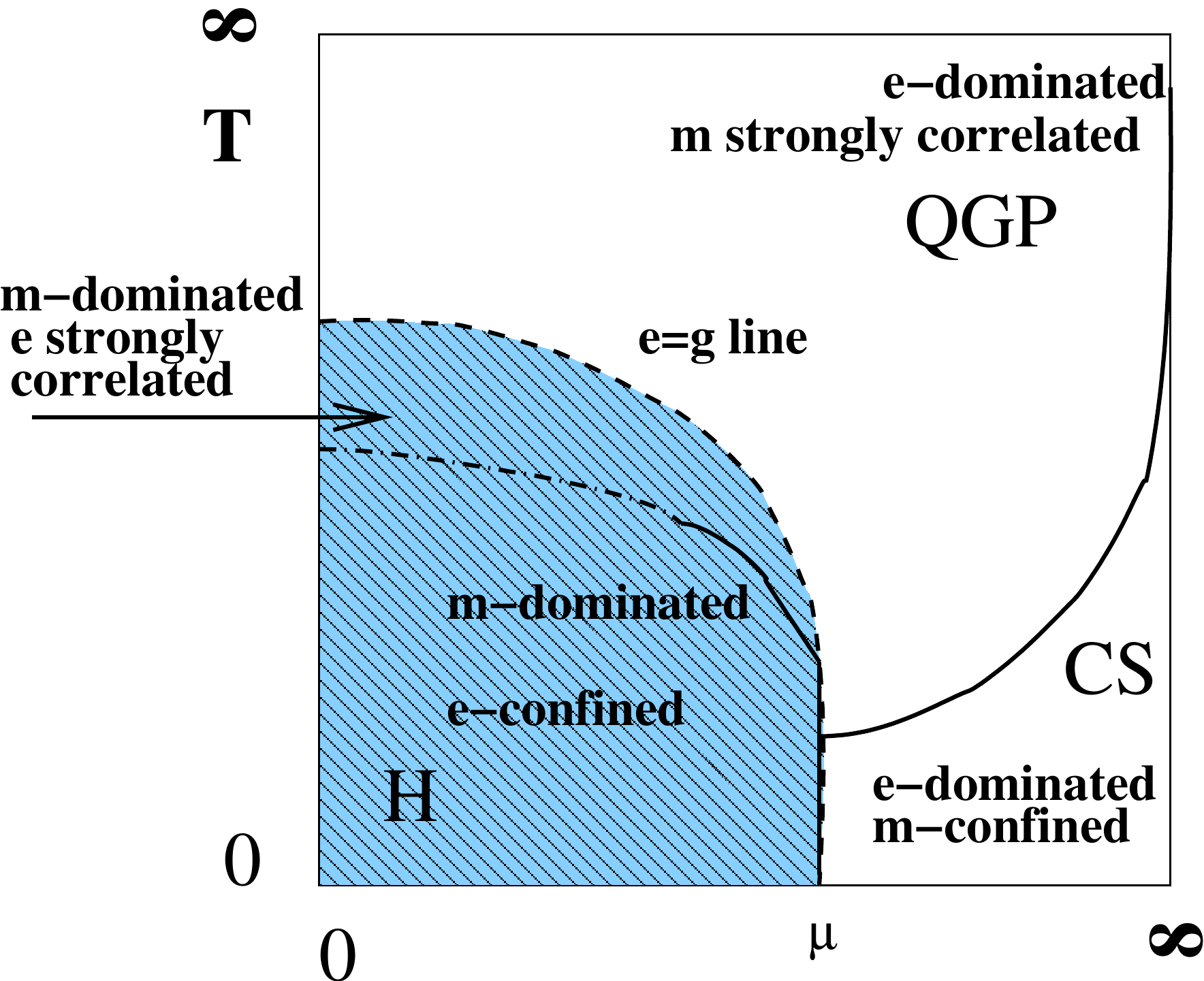}
 \includegraphics[width=0.4\textwidth]{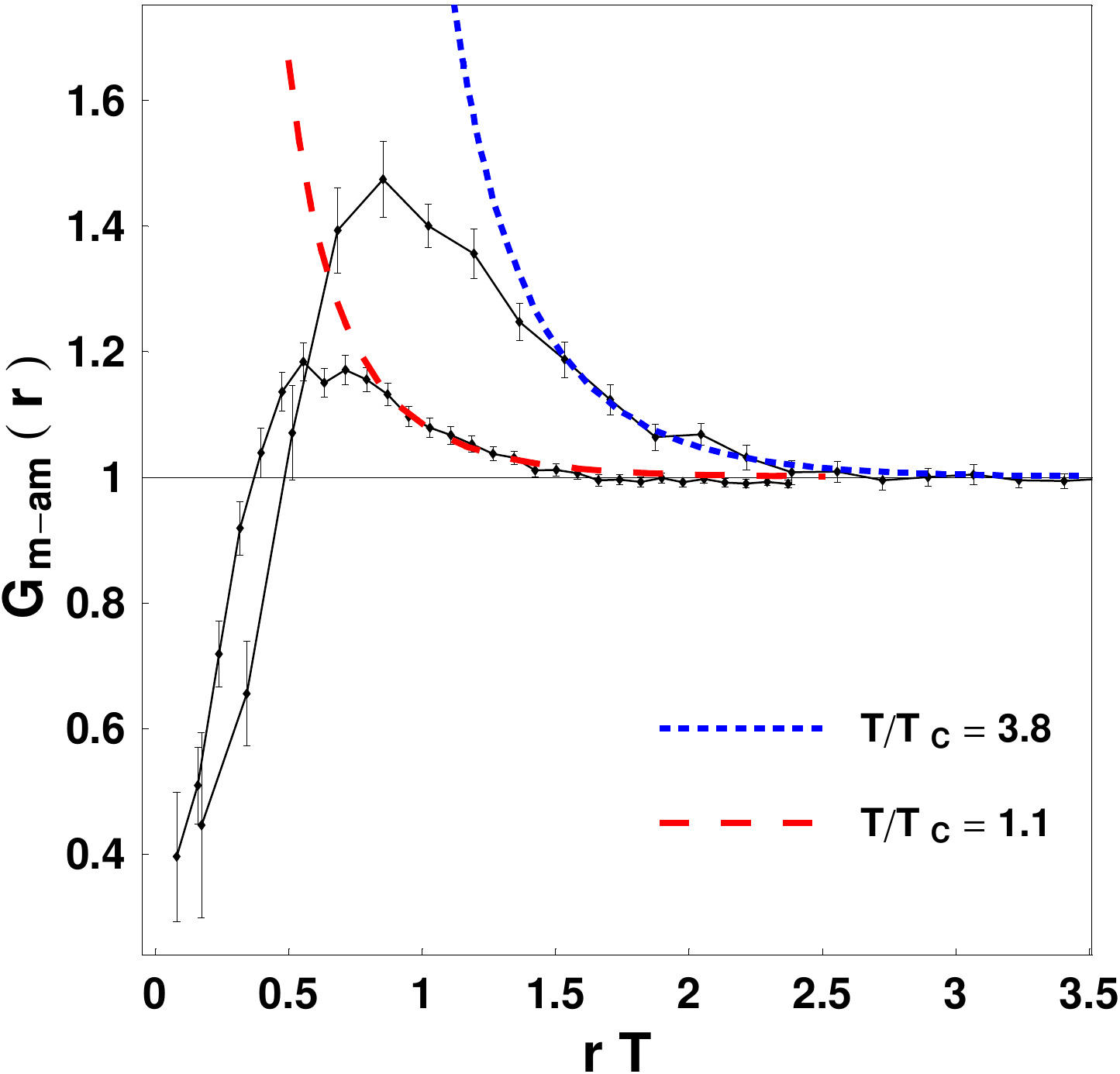}\\
  \includegraphics[width=0.48\textwidth]{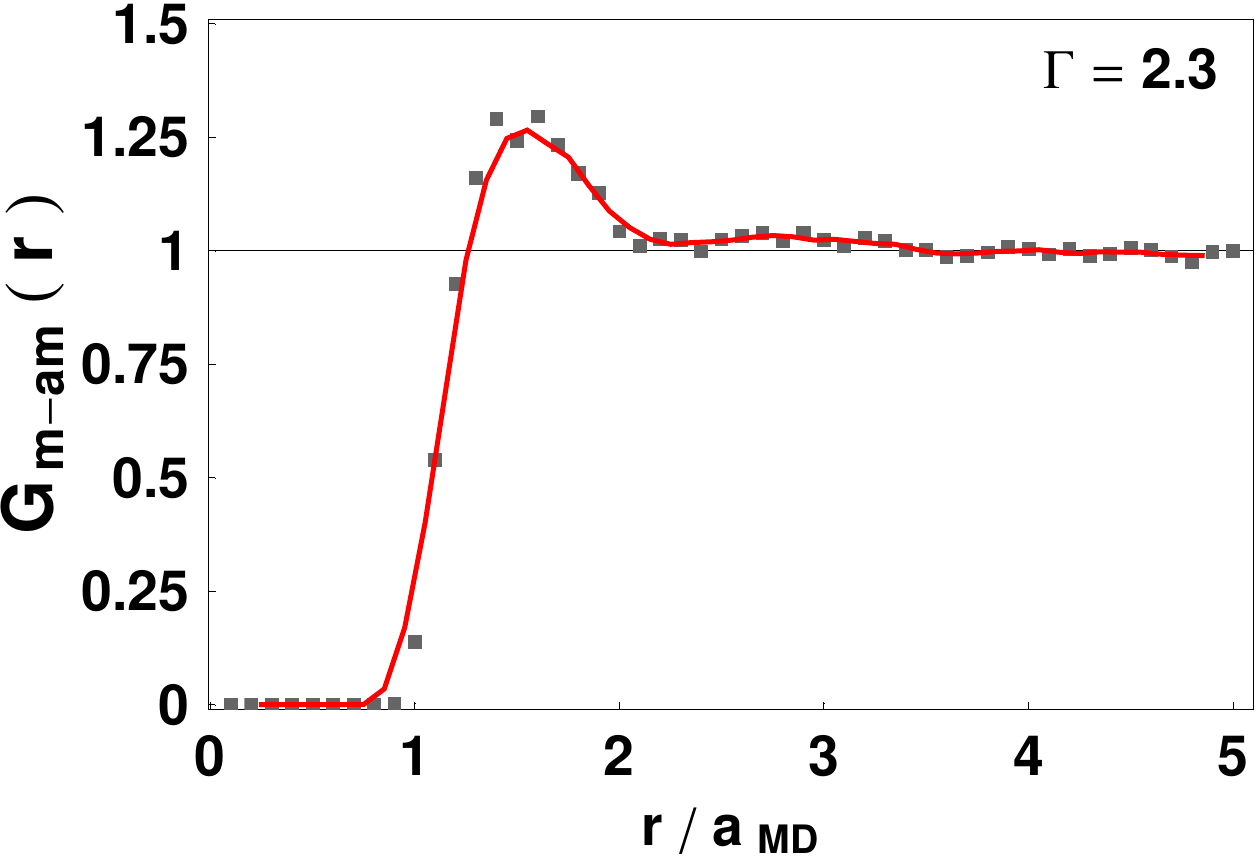}
 \includegraphics[width=0.45\textwidth]{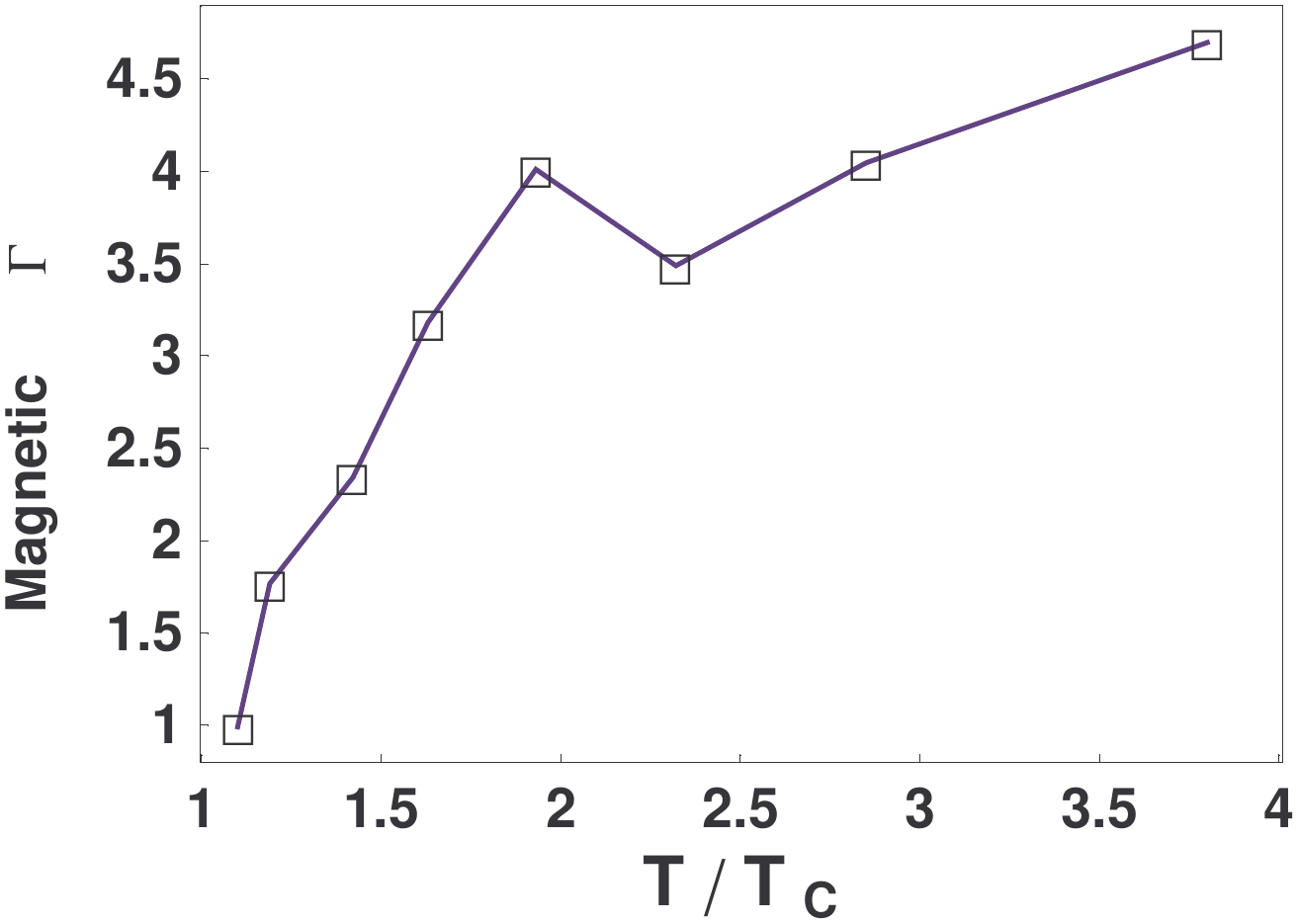}
 \vspace{0.1in}
\caption{(a)from \protect\cite{Liao:2006ry}:
 Schematic phase diagram on
 a (``compactified'') plane of
 temperature and baryonic chemical potential $T-\mu$.
 The (blue) shaded
region is ``magnetically dominated'', $g<e$, which
includes the e-confined hadronic phase as well as ``postconfined''
part of the QGP domain. Light region includes ``electrically
dominated'' part of QGP and also color superconductivity (CS)
region, which has e-charged diquark condensates and therefore
obviously m-confined. The dashed line called ``e=g \,line'' indicate
 electric-magnetic equilibrium.
 The solid lines show  phase transitions,
while the dash-dotted line is a deconfinement cross-over line.\\
(b)Monopole-antimonopole correlators
versus distance: points are 
lattice data \protect\cite{D'Alessandro:2007su}  for SU(2) pure gauge
 theory,
for the lowest and highest
temperatures, $T=1.1T_c$ and $T=3.8T_c$. The dashed lines 
are our fits from which magnetic couplings are extracted.  \\ 
(c)
 Monopole-antimonopole correlator, from our MD simulations 
\protect\cite{Liao:2008jg}.\\
(d)Effective magnetic plasma parameter $\Gamma_M$ 
 at various temperatures.\protect\cite{Liao:2008jg}\\
}
\label{fig_em_phasediag}
\end{figure}

As Dirac famously shown, quantum mechanics demands that
 electric and magnetic
coupling constants are related by the
 celebrated  quantization condition, which in quantum field theory
setting require them to run in the $opposite$
directions:
\begin{equation} \label{Dirac_quantization}
\alpha(electric)\alpha(magnetic)=1 \hspace{1cm} \beta(electric)+\beta(magnetic)=0
\end{equation}
Thus
 when $\alpha(electric)=e^2/4\pi$ is small
(at high T),  $\alpha(magnetic)=g^2/4\pi$ should be strong.
As $T$ decreases toward $T_c$, electric one  decrease and magnetic one
grows, till monopoles take over quarks and gluons: see
 schematic phase diagram
 shown in Fig.\ref{fig_em_phasediag}(a).
Recent lattice data \cite{D'Alessandro:2007su} provided dramatic
 conformation of this scenario.  Fig.\ref{fig_em_phasediag}(b)
shows two sets of these data, and the correlation 
(and thus magnetic coupling) is indeed stronger at $higher$ $T$.
Furthermore, the correlation function for 50-50 mix of
 electric/magnetic plasma obtained in our Molecular Dynamics (MD)
simulation  Fig.\ref{fig_em_phasediag}(c)
has the same shape and magnitude, provided one compare
at the same value of the magnetic plasma parameter
 $\Gamma \equiv \alpha(magnetic)  /  (\frac{3}{4\pi n})^{1/3}/{T}$:
its extracted values are shown in
 Fig.\ref{fig_em_phasediag}(d). It is very nice to find 
always $\Gamma>1$, which means that magnetic component of sQGP is also
liquid not gas, thus it does not spoil the ``perfect
liquid'' at RHIC. One may 
further think that viscosity has a minimum where both
electric quasiparticles (quarks) and magnetically ones
(monopoles) have similar difficulty propagating. We infer from
lattice data  that such {\em electric-magnetic equilibrium}
is at
$T\approx 1.5T_c$, right in middle of the RHIC domain.

 Transport properties for novel types of plasmas, including
electric and magnetic charges, have been calculated
by Liao and myself
 \cite{Liao:2006ry}: and $\eta$ is indeed minimal for most symmetric
 mixture  50-50\%. Before we turn to these results, let me 
qualitatively explain
 why in this case the diffusion/viscosity is maximally
reduced. Imagine
one of the particles - e.g. a quark. The Lorentz force makes it
rotate around a magnetic field line, which  brings it toward
one of the nearest monopoles. 
Bouncing from it, quark will go along the line to an antimonopole,
and then bounce back again: like electrons/ions do in the so called
``magnetic bottle''~\footnote{By the way, invented in 1950's by one of my teachers
G.Budker.}. Thus in 50-50 mixture all
 particles can be trapped between their dual neighbors, 
so that the medium can only
expand/flow collectively.

\begin{figure}
\centering
    \includegraphics[keepaspectratio,width=0.5\textwidth]{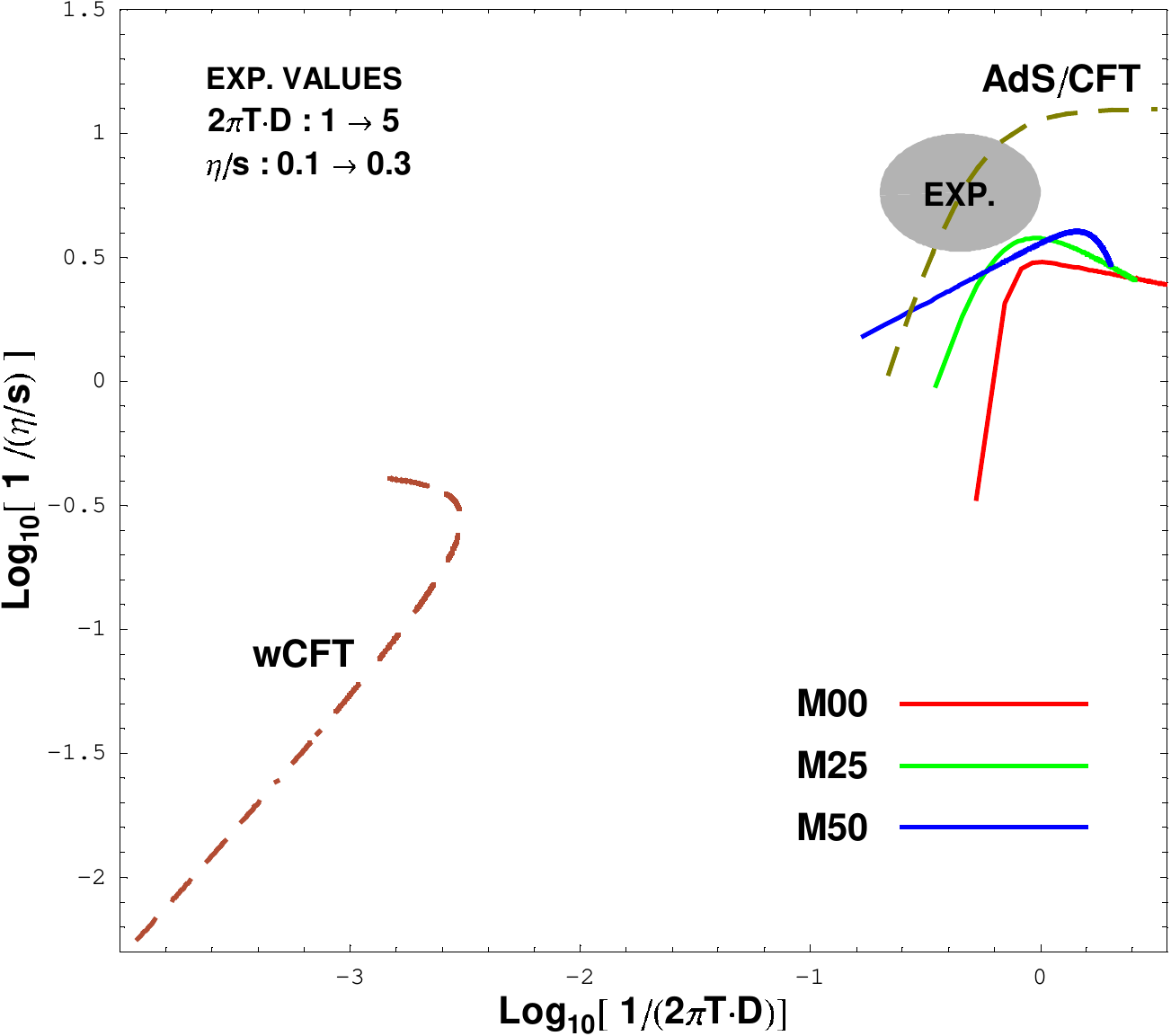}
 \caption[a]{Transport summary from \protect\cite{Liao:2006ry}:
 $Log[1/(\eta /s)]$ v.s. $Log[1/(2\pi T D)]$ including
results from our MD simulations, the AdS/CFT calculations, the weakly coupled
CFT calculations, as compared with experimental values, see text. }
\label{fig_vis_d_mapping}
\end{figure}

Our MD results  are shown on viscosity-diffusion plane in
Fig.\ref{fig_vis_d_mapping} by three lines: they are
compared to those from the AdS/CFT correspondence
in weak and strong coupling as well as 
with empirical values
 from RHIC experiments ( gray oval).
The dashed curve in the left lower corner is for $\cal N$=4 SUSY
YM theory in weak coupling:
 both quantities are proportional to the
same mean free path. These weak coupling results
are quite far from empirical data from RHIC
in the right upper corner. (Viscosity estimates follow from
deviations of the elliptic flow at large $p_t$ from hydro
predictions 
and diffusion constants are
estimated from $R_{AA}$ and elliptic flow of charm
.)
The  strong-coupling AdS/CFT results (viscosity according
to \cite{Policastro:2001yc} with $O(\lambda^{-3/2})$ correction, 
diffusion
constant from \cite{Casalderrey-Solana:2006rq})
are represented  by the upper dashed
line,  going right through the empirical
region. Our MD results -- three solid lines on the right --  
 are close to the
experiment as well, especially the 
version with the equal mixture of
EQPs and MQPs.

The last point I would like to make is
the {\em electric-magnetic competition} mentioned above.
An electric charge entering a region  with  magnetic
 field makes Larmor semicircle and
gets reflected back. Thus electric plasma (or Bose condensate)
is trying
 to expel magnetic field into
flux tubes. We know how this works  in superconductors
or in (e.g. solar) plasmas. Dual to that: magnetic
plasma expels the electric field. It does  happen not only in
a condensate (dual superconductor) phase at $T<T_c$, but in a 
QGP phase as well, under conditions derived in \cite{Liao:2007mj}.  
See Liao's talk here which explains how this phenomenon
explains unusual behavior of the heavy quark potentials 
at $T\approx T_c$.

\section{ AdS/CFT duality: conformal plasma in  equilibrium}
\begin{figure}[t]
\centerline{\includegraphics[width=12cm]{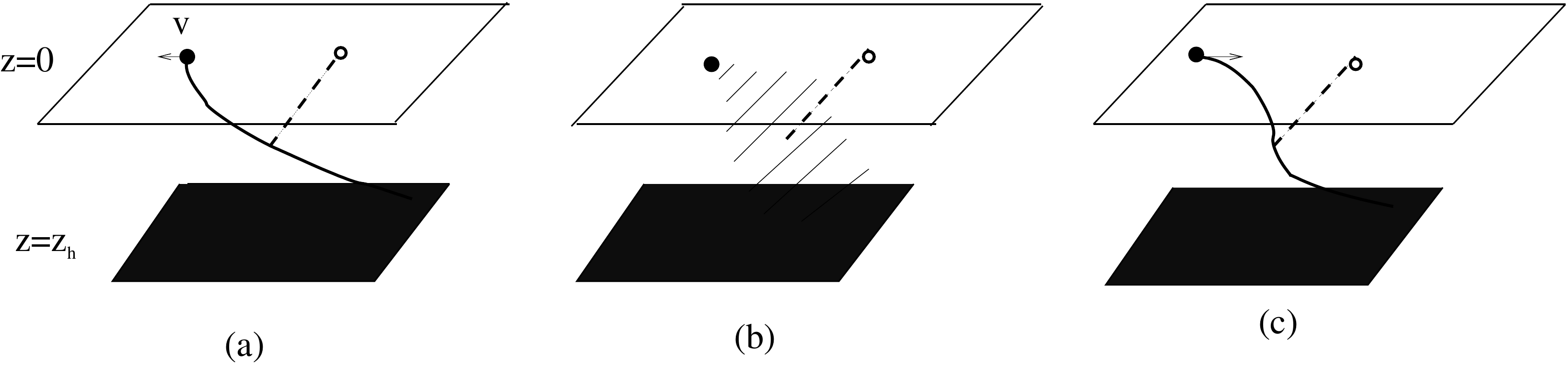}}
\caption{\small Schematic view of the relaxation settings,
a string (a), a wave (b) or a particle (c)
fall into the 5-th dimension toward the
black hole. 
} \label{fig_relaxation}
\end{figure}

Relation between RHIC physics and string theory was
already discussed at the previous QMs. Instead of an introduction 
(Rajagopal's talk here may have some),  
I will  provide some ``intuitive picture'' for non-experts.
Let me start with the finite-T Witten's settings
in which most\footnote{The exception is heavy quark diffusion
  constant  calculated by Casalderrey and Teaney\cite{Casalderrey-Solana:2006rq} which
 needs more complicated
settings, with a
 Kruskal metric connecting a World to an Anti-world through
the black hole.
}  pertinent calculations are done, shown in 
Fig.\ref{fig_relaxation}. The upper rectangle is the 
3-dimensional space boundary z=0 (only 2 dim shown),
 which is flat (Minkowskian) and corresponds
to ``our world'' where the gauge theory lives.
  Lower black  rectangles
(reduced in area because of curvature)  is corresponding patch
of the horizon
(at $z=z_h$) of a black hole whose center
is  located at $z=\infty$. Studies of
conformal plasma famously started from  
evaluation of the  Bekenstein entropy \cite{thermo}, 
$S=A/4$ with $A$ being the black patch $area$. 

(For non-experts: this setting can be seen as a swimming pool, with
 our gauge theory living on its surface, $z=0$, at
the desired temperature $T$. While pool's bottom 
is $infinitely$ hot,
strong gravity  stabilizes
 this setting, even thermodynamically.)

Fig.\ref{fig_relaxation}(a) shows a 
setting of heavy quark quenching \cite{quenching}:
a quark is  being dragged 
(at some hight $z_m$ related to the quark mass)
by an ``invisible hand'' (to the left): its electric flux goes into
 the 5-th dimension, into the so called
``trailing string''.  Its weight forces it to fall to the bottom
(horizon). (Think of 
a heavy quark  as a ship diligently laying
underwater cable to the pool's bottom.) The cost of that is the drag 
$$ {dP/dt}= -\pi T^2\sqrt{g^2 N_c} {v/2\over \sqrt{1-v^2}}
$$
connected to the diffusion constant  via
Einstein relation, a nontrivial  successful 
check on two very different calculations.

 Another form of relaxation is studied via propagating ``bulk
waves'' (b):  massless ones may have spin
 S=0 (dilaton/axion),1(vector) or 2 (gravitons).
Absorptive boundary condition at
the horizon (black bottom)leads to spectra of
``quasinormal\footnote{Quasinormal modes are those which
do not conserve the norm of the wave: it is like decaying
radioactive states in nuclear physics which are
distinct from scattering ones, with real energies. }
 modes''
with the imaginary part $Im(\omega_n) \sim \pi T n$,
setting the dissipation timescale of various fluctuations.
An exceptional case
  S=2 has two near-zero
modes, corresponding to only two propagating ``surface waves'',
the longitudinal sound and transverse ``diffuson''.
Absorption at
the bottom (horizon) of both
 famously gives the viscosity $\eta/s=1/4\pi$\cite{Policastro:2001yc}.
The waves may have real (timelike) 4-momentum or virtual
(spacelike) one\footnote{
 This  case, named DIS in AdS,  is
 discussed  here   by E.Iancu.}. Rather 
complete spectra of quasinormal
 modes and spectral densities for S=0,1,2 correlators are  
available, unfortunately extracted numerically.
 The case (c) -- a ``falling stone'' --  perhaps represent
colorless (no strings attached)
 ``mesons'', released to plasma and relaxing. 

 
The dashed lines in Fig.\ref{fig_relaxation} corresponds to
the next-order diagram, describing back reaction
of the falling bulk objects onto the boundary,
the observation point denoted by a small open circle. This fields 
may also have
spins 0,1 or 2, providing 3 pictures known as
a (4-d) ``holograms'' of the bulk. Contrary to our intuition (developed from
our limited flat-world experience),  the hologram is $not$   
a reduced reflection of more complete 5-d dynamics in the bulk, but
in fact represents it fully. This phenomenon
 -- the AdS/CFT duality -- is a miracle occurring due to near-black-hole
setting.

\begin{figure}[t]
\includegraphics[width=0.3\textwidth]{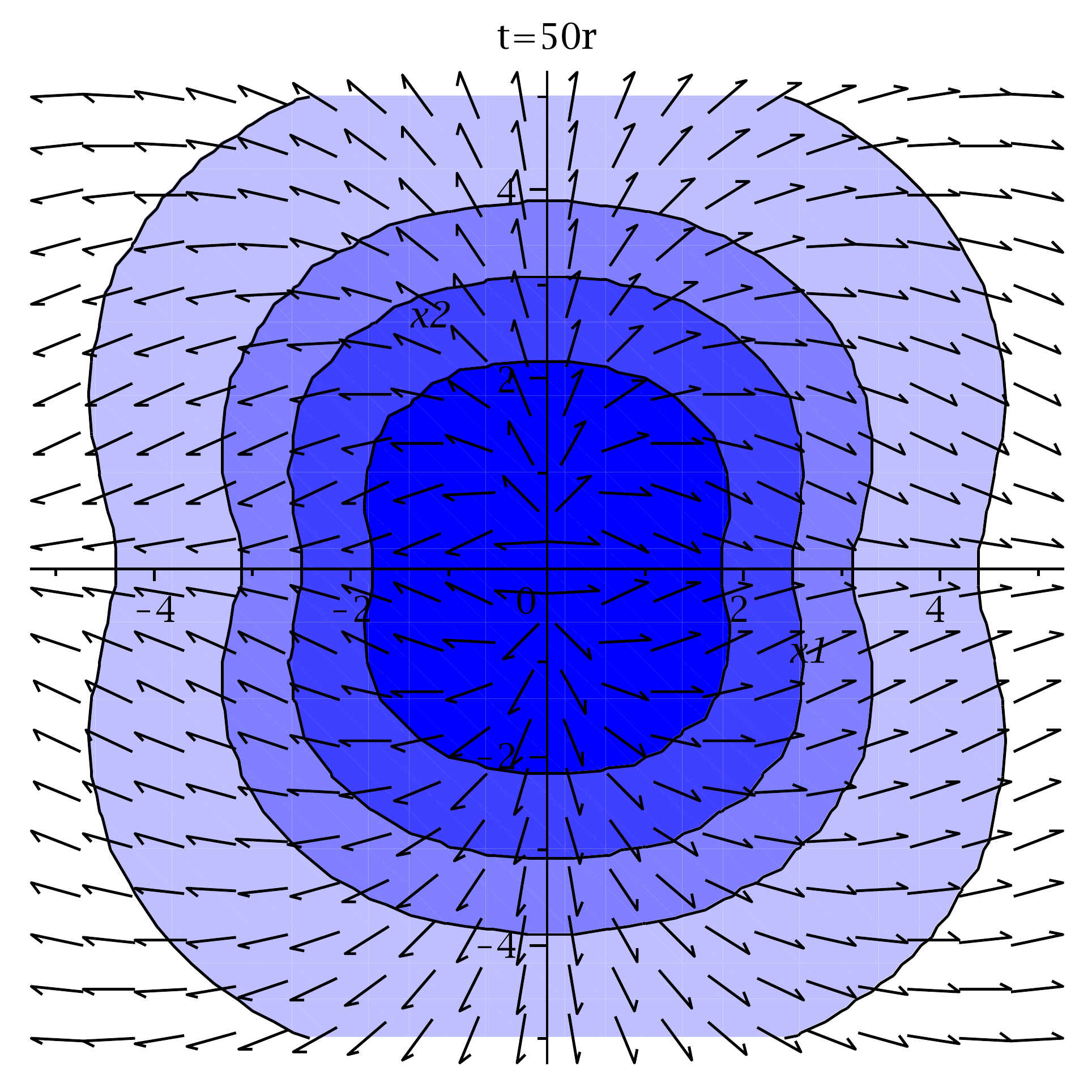}
\includegraphics[width=0.7\textwidth]{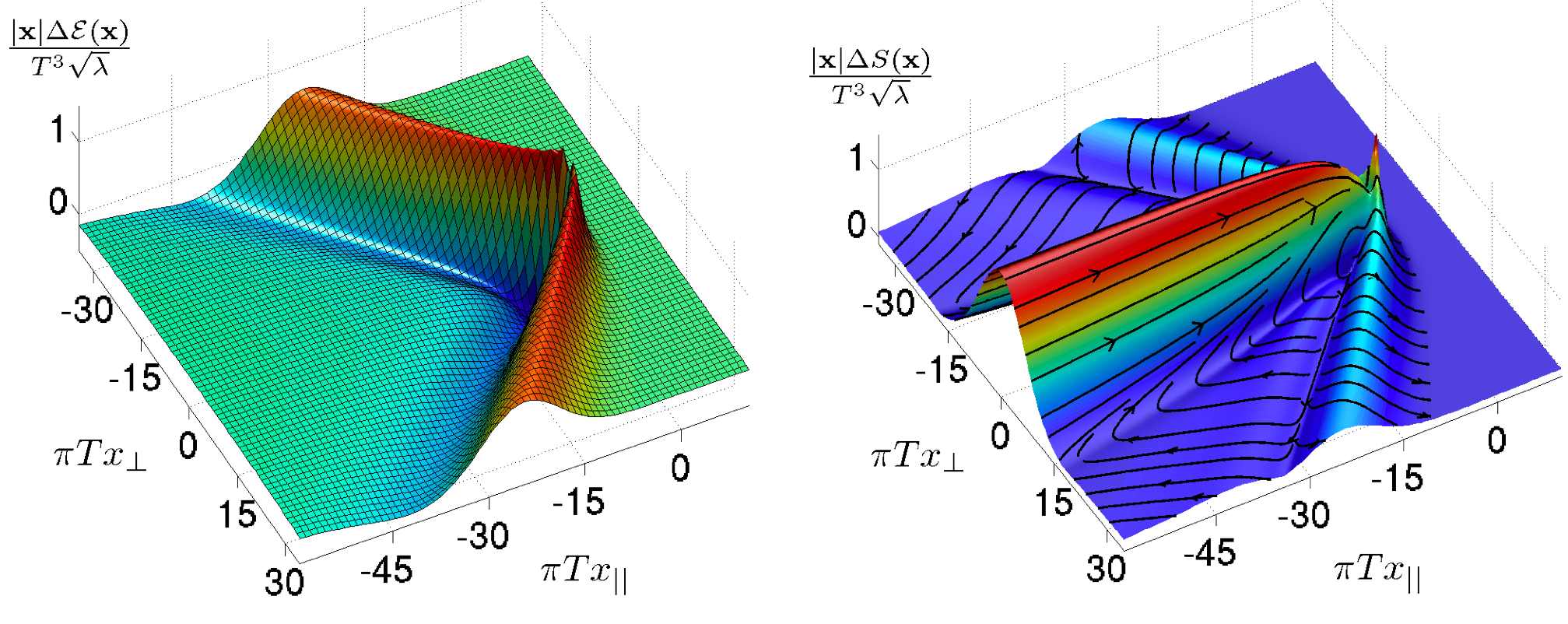}
\caption{\label{fig_holog}
(a) From \protect\cite{Lin:2006rf}: The hologram of a falling string.
The contours show the magnitude of the Poynting vector
$T^{0i}$ in the transverse plane,
The direction of the momentum flow is indicated by arrows.
(b) From \protect\cite{Chesler:2007sv}: 
hologram of the trailing string,
the normalized energy density 
 for one quark (supersonic jet) with $v = 3/4$ at nonzero $T$.
(c) same as (b) for the Poynting vector.}
\label{fig_holo}
\end{figure}

These holographic images are what the surface observer will
see.  Image of the trailing string was calculated in \cite{conical_ads,Chesler:2007sv}:
 the recent example at nonzero $T$
is shown in Fig.\ref{fig_holog}(b,c): it accurately
displays hydro  conical flow. For a hologram of the  stone Fig.\ref{fig_relaxation}(c)
see recent paper \cite{Gubser:2008vz}: but to our knowledge
the holographic ``back reaction''
 of the falling waves remains to be done.

 How these predictions are related to experiment? 
Apart of those
shown in Fig.\ref{fig_vis_d_mapping},  important
 test is whether
the  drag force 
indeed depends only on the  $velocity$ (rather 
than momentum): can be done via single electrons from $c$ and $b$ 
decays. 
Another challenge is to test
if the effective
viscosity is indeed $decreasing$  with increasing gradients 
as AdS/CFT nontrivially indicate \cite{Lublinsky:2007mm}:
 it can 
be inferred from elliptic flow at more peripheral collisions
(thinner ``almond''). 

\section{ AdS/CFT duality: equilibration and sGLASMA}
New challenging
 frontier is AdS/CFT {\em  out of equilibrium}, addressing
 initial equilibration  and entropy production.
As explained in Venugopalan's talks, 
{\em ``glasma''}  is a
non-equilibrium gluonic state between the
collision moment and equilibrated QGP, which  so far 
is modeled by
random  glue 
via classical Yang-Mills eqn in $weak$ coupling. However 
the corresponding 
``saturation scale'' $Q_s$ at RHIC is only about 1-1.5 GeV -- not far
from parton momenta in sQGP, the perfect liquid as one knows -- so one
may wander if a $strongly$ coupled regime should be
 tried instead.

 This is what I propose to call sGLASMA frontier:
 AdS/CFT is the tool to use. It means that one has to start with
 high energy collision inside cold $T=0$ $AdS_5$ (the vacuum,
or a $bottomless$ pool) and then dynamically
solve two difficult problems: (i) explain why ``
collision debris'' may act like 
a ``heater'' imitating black/hot patch of Fig.\ref{fig_relaxation};
(ii) find a consistent solution with  ``falling bottom'', $z_h(time)$,
and find its
hologram describing hydro explosion/cooling. 

\begin{figure}[t]
\centerline{\includegraphics[width=12cm]{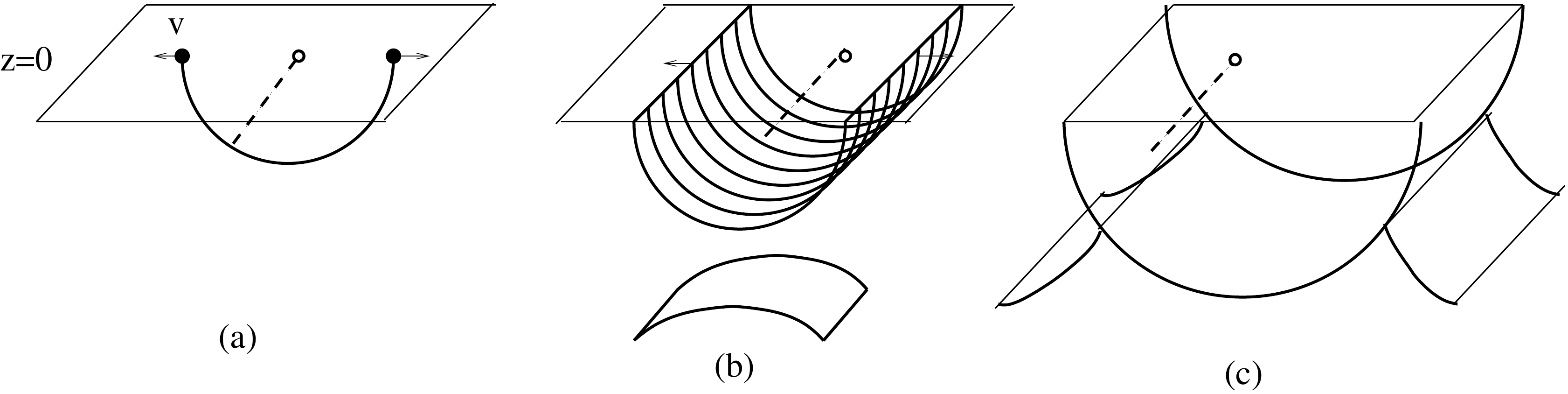}}
\caption{\small Schematic view of the collision setting.
Setting of the sGLASMA studies: (a) a single pair of heavy quark
jets,
moving with velocities $v$ and $-v$ and creating falling string.
Multiple strings create a 3-d falling membrane (2d shown),
which is (b) first far from trapped surface and then
very closed to it (c). 
} \label{fig_membranes}
\end{figure}

 Example of recent progress on the former (i) front are
works by S.Lin and myself \cite{Lin:2006rf} and more recently by
Hofman and Maldacena\cite{Hofman:2008ar}. One may view them as
steps toward a ``strongly coupled collider physics'',
with a $single$ pair of heavy quarks jets
 produced. 
Like in  Lund model (Pythia),
they are connected by a flux tube (string), which
 is however not  breaking but rather 
falling into the 5-th $z$ direction, Fig.\ref{fig_membranes}(a).
For one string one can  both
  solve eqns of falling and then find its (gravitational) hologram
\cite{Lin:2006rf}. The result is an explosion shown in
 fig.\ref{fig_holo}(a), which is however  {\em non-thermal and
thus non-hydrodynamical}.

 Temperature/entropy only appear when
a horizon (also called ``trapped surface'') is dynamically created 
leading to the information loss. A lot of work was done on
gravitational collapse, there are black holes in the Universe
and, with modified multidimensional gravity, 
people are thinking about their possible formation in  LHC
experiments. However, in AdS/CFT language we are sure that
$each$  RHIC heavy ion collision event does produce a black hole, but
with an effective  gravity (imitating QCD) in the imaginary
(unreal) 5-th dimension. In heavy ion context, 
 Sin, Zahed and myself~\cite{Shuryak:2005ia}
first argued that exploding/cooling
fireball on the brane is dual to
 $departing$  black hole,  formed by the collision debris and then 
falling toward the AdS center. A specific solution they discussed in the paper 
was a brane departing from a static black hole, which generated
a ``spherical'' solution
(no dependence on all 3 spatial coordinates) with a time-dependent
$T$ 
(which however is more 
 appropriate for cosmology but not heavy ion
applications). These authors also discussed other idealized settings, 
with d-dimensional stretching, corresponding for d=1 to
a collision of two infinite  thin walls and subsequent Bjorken
rapidity-independent expansion, with 
 2d and 3d corresponding to cylindrical and spherical 
relativistic collapsing walls.

 Instead of solving Einstein equations with certain
source, describing
gravitationally collapsing ``debris'' of the collision,
Janik and Peschanski \cite{Janik:2005zt}
applied an ``inverse logic'',  extrapolating
  into the bulk the metric which yield
 expected hydrodynamical solution at the boundary.
They found asymptotic (late-time) solution corresponding to
 1+1-dim rapidity-independent Bjorken expansion. It indeed has   
 a departing $horizon$ at $z_h\sim \tau^{1/3}$.
 Important feature of this leading order solution
is that while the horizon is stretching in one 
direction it is contracting in others keeping the
  total horizon  area constant: this is entropy
conservation.
 The first subleading terms $O(\tau^{-2/3})$ 
has been calculated by Sin and Nakamura  \cite{Nakamura:2006ih}  who 
 identified them with the viscosity effects, although the
viscosity value was only fixed by still further term by
 Janik et al. However  they eventually
 concluded \cite{Benincasa:2007tp} that
the expansion series are inconsistent beyond the first few orders.
 I always argued this should be the case:
 a near-horizon singularity which they see as a
problem just shows 
inevitability of the matter presence: pure gravity 
simply is not enough.

Further work toward working out a  ``gravity dual''
to heavy ion fireball is ongoing: let me show just a sketch
of our current work.
 If many strings are falling together their
 combined gravity is non-negligible -- they are partly folling
under their own weight. So
one should solve non--linearized Einstein eqns, which
tell us that (from the viewpoint of distant observer)
extra weight may actually slow down falling, eventually
leading to near-horizon
levitation. 
The trapped surface  is moving first upward (shown at the bottom
of   Fig.\ref{fig_membranes}(b))
 toward the falling
membrane, till two collide, get close and fall together, see 
Fig.\ref{fig_membranes}(c). After that distant observer finds
 a thermal hydrodynamical explosion as a hologram. This is the case
 at mid-rapidity
 but never
in the fragmentation regions.

Finally, let me mention a separate direction by Kajantie
et al \cite{Kajantie:2007bn} addressing
these issues in the 1+1 dimensional world. It is easier to work out
math in this case: but shear viscosity is absent in it and
bulk viscosity is prohibited by conformity, so it is a cute
toy case without dissipation.

\end{document}